\documentclass[runningheads]{llncs}
\usepackage[T1]{fontenc}
\usepackage{graphicx}

\usepackage{amsmath}
\usepackage{amssymb}
\usepackage{xcolor}
\usepackage{enumitem}
\usepackage{tabularx}
\usepackage{booktabs}  
\usepackage{array}
\usepackage{hyperref}
\usepackage{multirow}
\begin{document}
\title{
Diffusion as Sound Propagation: Physics-inspired Model for Ultrasound Image Generation}
%
\titlerunning{Physics-inspired Model for Ultrasound Image Generation}
%

\author{Marina Domínguez\inst{1,2 *} \and
Yordanka Velikova\inst{1,2 *} \and
Nassir Navab\inst{1,2} \and
Mohammad Farid Azampour\inst{1,2,3}
}

\authorrunning{Domínguez et al.}
\institute{Computer Aided Medical Procedures, Technical University of Munich, Germany \\ 
 \and
Munich Center for Machine Learning, Munich, Germany \\
\and
 Department of Electrical Engineering, Sharif University of Technology, Tehran, Iran \\ 
 }
\maketitle              

{\let\thefootnote\relax\footnotetext{*Shared first authorship.}} 
\begin{abstract}
Deep learning (DL) methods typically require large datasets to effectively learn data distributions. However, in the medical field, data is often limited in quantity, and acquiring labeled data can be costly. To mitigate this data scarcity, data augmentation techniques are commonly employed. Among these techniques, generative models play a pivotal role in expanding datasets. However, when it comes to ultrasound (US) imaging, the authenticity of generated data often diminishes due to the oversight of ultrasound physics. 

We propose a novel approach to improve the quality of generated US images by introducing a physics-based diffusion model that is specifically designed for this image modality. The proposed model incorporates an US-specific scheduler scheme that mimics the natural behavior of sound wave propagation in ultrasound imaging.
Our analysis demonstrates how the proposed method aids in modeling the attenuation dynamics in US imaging.
We present both qualitative and quantitative results based on standard generative model metrics, showing that our proposed method results in overall more plausible images. Our code is available at \href{https://github.com/marinadominguez/diffusion-for-us-images}{github.com/marinadominguez/diffusion-for-us-images}.

\end{abstract}

\section{Introduction}
The scarcity of labeled medical data poses a significant challenge for training deep learning models, thereby encouraging the exploration of alternative solutions.
Generative models have emerged as a popular approach to address this issue, allowing the generation of synthetic data that complements the limited available labeled examples \cite{Kebaili2023review}.
By producing synthetic samples, generative models, such as diffusion models, can significantly reduce the problem of data scarcity, protect patient privacy, and address class imbalance \cite{Kazerouni2022DiffusionModels}.

Contrasting to standard camera images, ultrasound's image formation process relies on the interpretation of echo patterns, which necessitates specialized approaches for generating realistic synthetic.
B-mode ultrasound formation process involves emitting ultrasound pulses and capturing returning echoes generated by the interaction of sound waves with tissues
\cite{Szabo2021EssentialsUltrasound}. 
This interaction leads to reflection, refraction, and attenuation, posing challenges in capturing internal structures \cite{Grogan2023UltrasoundPhysics}. Understanding these challenges, particularly attenuation, is crucial, as the upper regions appear more defined and brighter due to the stronger signal, while the lower regions become darker as the signal diminishes with depth. Addressing this gradual attenuation is key to enhancing the realism of synthetic ultrasound images.

Diffusion models in US imaging were initially used for tasks such as denoising and image generation. Initial studies have focused on reducing speckle noise and improving image clarity \cite{Goudarzi2023DeepUltrasound,Zhang2022UltrasoundDenoising,Schaft2021UltrasoundSpeckle}. More recent research propose methods that reduce noise but preserve the speckle texture, enhancing image quality \cite{Asgariandehkordi2023DUDP_newUS}. Experimental results from these studies show that such methods outperform traditional denoising techniques in both Peak Signal to Noise Ratio (PSNR) and Generalized Contrast to Noise Ratio (GCNR) \cite{RodriguezMolares2020GCNR}. 
Currently, the use of these models extends beyond denoising tasks. This includes semi-supervised learning for US segmentation \cite{Tang2023MLGCC_newUS} and image generation from semantic maps~\cite{stojanovskiEchoFromNoise2022}, both works showcasing significant improvements in segmentation accuracy. These studies demonstrate the power of diffusion models in enhancing US imaging and the potential for boosting DL models in tasks like image segmentation~\cite{misegnet}.

While diffusion models have shown remarkable success in generating high-quality images across various domains, their direct application to US imaging overlooks its physical properties critical to this modality~\cite{tirindelli2021rethinking,cactuss,lotus}. Considering the significant differences between B-Mode ultrasound and natural images, applying the same synthetic image generation methods to both is impractical. Consequently, we have developed an approach that adapts standard diffusion models to better align with the actual process of US image generation.

\subsubsection{Contributions}
This paper presents a novel approach to diffusion models designed specifically for US image generation. We propose a new noise scheduler inspired by the natural behavior of sound wave propagation. 
This scheduler simulates the attenuation of echoes returning to a US receiver. 
We consider the changes in depth-dependent US resolution and put more emphasis on regions closer to the probe, where images inherently show greater clarity, detail, and reliability of internal structures. 
We evaluate and compare the generated images qualitatively and quantitatively against a baseline with a conventional noise scheduler, both with and without semantic labels. 

\section{Methodology}
This section details the adaption of diffusion models for US synthesis, by integrating a novel noise scheduler: the B-maps. We show that by introducing this scheduler, which is designed to mimic the natural attenuation of sound waves interacting with tissues, we are able to generate more plausible B-mode images. 

\subsection{Background}

\subsubsection{Forward process.}
DDPM \cite{hoDenoisingDiffusionProbabilisticModels2020} defines the forward diffusion process as a Markov chain where Gaussian noise is added in successive steps to obtain a set of noisy samples. Consider $q(x_0)$ as the uncorrupted (original) data distribution. Given a data sample $x_0 \sim q(x_0) $, a forward noising process $p$ which produces latent $x_1$ through $x_T$ by adding Gaussian noise at time $t$ is defined as follows \cite{Asgariandehkordi2023DUDP_newUS}:
\begin{equation}
q(x_t | x_{t-1}) = \mathcal{N} (x_t; \sqrt{1 - \beta_t} \cdot x_{t-1}, \beta_t \cdot I), \quad \forall t \in \{1, \ldots, T\},
\end{equation}
where $T$ represent the number of diffusion steps and $ \beta_1, \ldots, \beta_T \in [0, 1] $  the noise scheduler across diffusion steps \cite{Kazerouni2022DiffusionModels}. Considering $\alpha_t = 1 - \beta_t$ and $\bar{\alpha}_t = \prod_{t=1}^{T} \alpha_t$, by applying the parametrization trick: $x_t = \sqrt{\alpha_t} x_0 + \sqrt{1 - \alpha_t} \varepsilon$, $t$ times, one can directly sample a  step of the noised latent conditioned on the input $x_0$ \cite{luoUnderstandingDiffusionModels2022}:
\begin{equation}
q(x_t | x_0) = \mathcal{N} (x_t; \sqrt{\bar{\alpha}_t}x_0, (1 - \bar{\alpha}_t) I)
\label{standard_forward}
\end{equation}
\subsubsection{Reverse process.} 
The reverse process seeks to approximate a sample from the original data distribution $q(x_0)$ by starting from a standard Gaussian distribution $p(x_T) = \mathcal{N} (x_T; 0, I)$ and iteratively denoising towards $x_0$. To this end, we can parameterize this reverse process as follows:

\begin{equation}
p_{\theta} (x_{0:T}) = p(x_T) \prod_{t=1}^{T} p_{\theta} (x_{t-1} | x_t) 
\end{equation}

\begin{equation}
p_{\theta} (x_{t-1} | x_t) = \mathcal{N} (x_{t-1}; \mu_{\theta} (x_t, t), \Sigma_{\theta} (t)),    
\label{standard_backward}
\end{equation}

employing learned parameters to guide the reverse diffusion towards accurate reconstruction of the original data. This dual-phase approach can create high-quality images from noise, laying a solid foundation for our method.

\subsection{B-Maps Definition}
\begin{figure}[t]
    \centering
    \includegraphics[width=\textwidth]{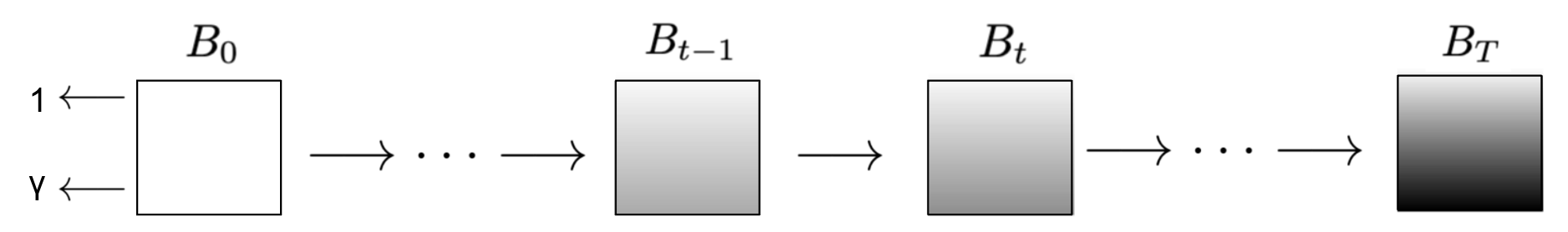}
    \caption{\textbf{Evolution of B-maps across time-steps.} In every timestep, the values in the B-Maps decrease top-to-bottom from 1 to a number, $\gamma$. As the timestep increases, $\gamma$ goes from 1 to $1 - \epsilon$, with $\epsilon$ being a small fixed value in the interval $(0,1)$.} 
    \label{fig:bmaps}
\end{figure}

Central to our approach is the introduction of B-Maps; these matrices, have the same dimensions as the US images, and allow for precise control of the noise level at each pixel. 
The idea is to change how the standard DDPMs introduce noise in the image. While the normal DDPM, originally designed for natural images, applies noise uniformly in the image, the B-Maps scale the noise across the vertical axis of the image, simulating the top-to-bottom image construction of US imaging. Our novel adaptation introduces more noise—and thus, faster convergence towards a standard Gaussian distribution—at the bottom of the image than at the top. As a result, we define a diffusion model that focuses on learning the distribution of the upper region of the image first before addressing the inherently noisier lower region, as the sound waves lose strength the deeper they go into the body. This prioritized learning process ensures that the most reliable details—those closer to the probe—are captured with higher fidelity and we also prevent coming up with artifacts and non-anatomically plausible features. 
An illustration of the definition of B-Maps is shown in Figure \ref{fig:bmaps}. 

\subsection{Denoising Diffusion Probabilistic Models with B-Maps}

\subsubsection{Forward Process.} 
\begin{figure}[b]
\includegraphics[width=\textwidth]{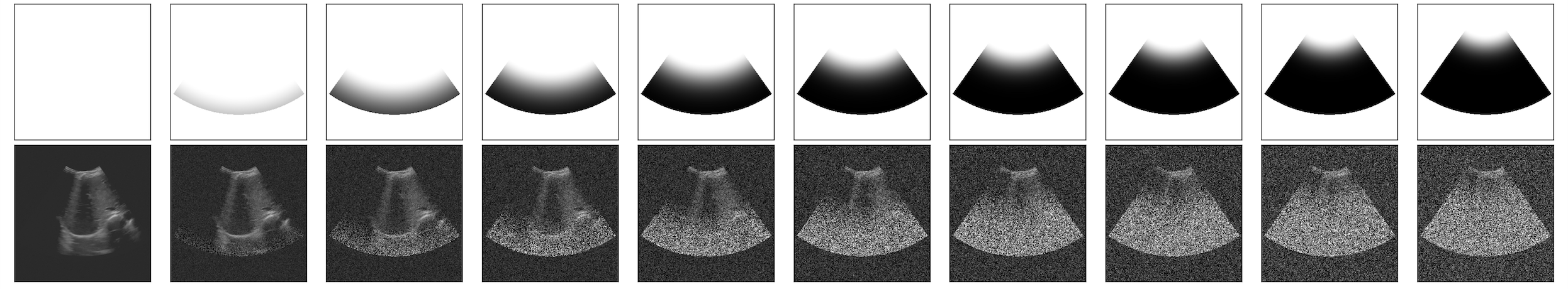}
\caption{\textbf{Forward pass:} Noise addition from bottom to top. Linearly-scheduled cone-shaped B-Maps on the top row and the visualization of the noising process of the US image in the bottom row. B-Maps are applied at each step, making the gaussian distribution converge earlier on the bottom than on the top.}
\label{fig:forward}
\end{figure}

We modify the standard forward pass (Eq. \ref{standard_forward}) that introduces the noise in the image centered in $ \bar{\alpha_t} x_0$ by integrating noise in a way that better reflects US image formation. By incorporating our B-maps noise scheduler, we specifically adjust the distribution's mean and variance towards the image's lower part. This adjustment is done by point-wise multiplying the existing noise schedule $\alpha_t$ and the preceding image $x_{t-1}$ with B-map scheduler $B_t$ at time $t$. This leads to our proposed new forward distribution: 

\begin{equation}
q(x_t | x_{t-1}) = \mathcal{N}(x_t; \sqrt{\bar{\alpha}_t \cdot \bar{B_t}} x_{t-1}, (1 - \bar{\alpha}_t \cdot \Bar{B_t}) \mathbf{I}) 
\label{eq:encoder_with_B_maps}
\end{equation}
with: $\bar{\alpha}_t = \prod_{t=1}^{T} \alpha_t \text{ and } \bar{B}_t = \prod_{t=1}^{T} B_t$ and $\cdot$ denoting the point-wise multiplication. This method guides the diffusion model's behavior, introducing noise in the forward process by simulating the progressive attenuation of echo intensity. 
The visual representation of this modified process is shown in Fig. \ref{fig:forward}.

\subsubsection{Reverse Process.} 
\begin{figure}[t]
    \centering
    \includegraphics[width=\textwidth]{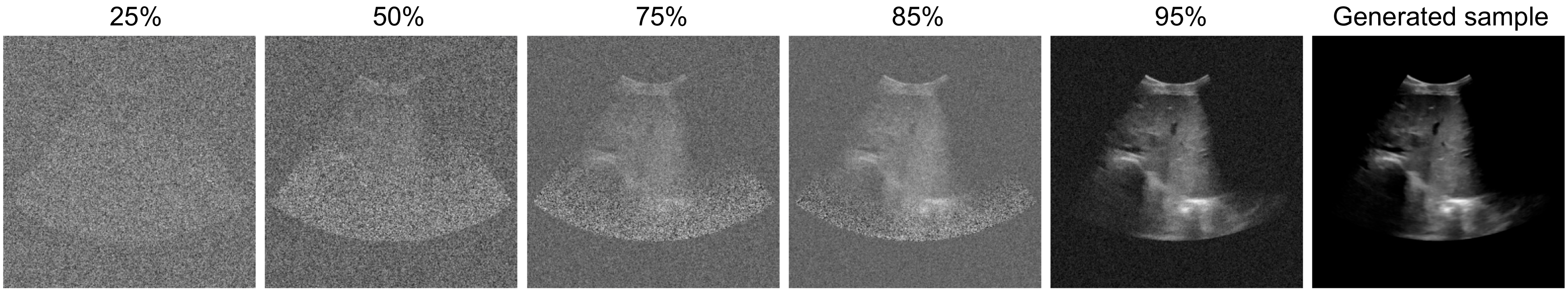}
    \caption{\textbf{Reverse Process:} denoising the image. Initially focusing on the area near the probe, the model progresses to denoise the image toward the bottom, mimicking the way US images are traditionally generated.}
    \label{fig:backward_with_b}
\end{figure}

Incorporating B-maps into our model alters traditional distributions (Eq. \ref{standard_backward}), leading to new derivations for the reverse process. These derivations can be found in the Supplementary Material (Sec \ref{sec:supp}). The newly derived posterior distribution for our physics-inspired diffusion model would be:
\[
q(x_{t-1}|x_t, x_0) \propto \mathcal{N}\left( x_{t-1}; \mu_\theta(x_t, t), \Sigma_\theta(t) \right),
\] where 
\[
\mu_\theta(x_t, t) = \frac{\sqrt{\alpha_t \cdot B_t (\mathbf{1} - \bar{\alpha}_{t-1} \cdot \Bar{B}_{t-1}
)} x_t + \sqrt{\bar{\alpha}_{t-1} \cdot \Bar{B}_{t-1}(\mathbf{1} - \alpha_t \cdot B_t)}x_0}{\mathbf{1} - \bar{\alpha}_t \cdot \Bar{B_t}}
\] and \[
\Sigma_\theta(t) = \frac{(\mathbf{1} - \alpha_t \cdot B_t)(\mathbf{1} - (\bar{\alpha}_{t-1} \cdot \Bar{B}_{t-1}))}{\mathbf{1} - (\bar{\alpha}_t \cdot \Bar{B}_t)}\mathbf{I}
\]
This reverse distribution considers the varying noise levels influenced by B-maps, creating a more natural top-to-bottom reconstruction of US images, as we can see in Figure \ref{fig:backward_with_b}.

In our work, we introduced this novel noise scheduler into Guided-Diffusion (GD) \cite{dhariwalDiffusionModelsBeat2021} and Semantic Diffusion Models (SDM) \cite{wangSemanticImageSynthesis2022}. Guided-Diffusion is based on the Improved DDPMs \cite{nicholImprovedDenoisingDiffusion2021} and enhances the generation process by utilizing guidance mechanisms. 
SDMs, on the other hand, extend the diffusion framework by incorporating semantic information.
SDMs include the integration and utilization of semantic labels during the generation process and greater control over the output characteristics by leveraging semantic information. 

We modified these models by replacing the original equations that define the diffusion process with our equations incorporating B-Maps. Specifically, the equation that introduces noise in the forward pass of these baselines was replaced with our derived Eq. (\ref{eq:encoder_with_B_maps}). Additionally, we defined our noise scheduler, the B-Maps, and adjusted the reverse process to incorporate them. Consequently, for the original variance in the models, we now use our newly derived variance, $\Sigma_\theta(t)$. These adaptations allow us to generate US images from datasets with and without semantic labels, using SDM and Guided-Diff correspondingly. By incorporating our B-Maps into these pre-existing frameworks, we can compare the effectiveness of our approach against the original models without B-Maps. 

\section{Experimental Setup}

\subsection{Datasets}

\subsubsection{SegThy}
dataset contains annotated 3D US images of the thyroid \cite{thyData} from 28 healthy volunteers, acquired with Siemens Acuson NX-3 US machine with a 12MHz VF12-4 probe. We extracted 2D slices and labels from the 3D US scans and removed the images without a thyroid label or with empty labels, totalling 2,250 images, where 512 were used for validation.

\subsubsection{CAMUS}
dataset includes 400 patient images for training and 50 for validation~\cite{camus}. Each patient contributes four images at both end-diastole (ED) and end-systole (ES) across two- and four-chamber views. In total, 1600 training and 200 validation images. Following Stojanovski et al. \cite{stojanovskiEchoFromNoise2022}, we applied five random affine and elastic deformations. This augmented the dataset to 8000 training and 1000 validation images.

\subsubsection{Liver}
images were acquired using a tracked probe ACUSON Juniper (Siemens Healthineers, Erlangen, Germany) with a 5C1 convex probe. We scanned 14 volunteers aged between 22 and 34. After excluding images with more than 50\% shadow, we ended up with 6,900 2D slices, where 1000 were used for validation. The images were horizontally padded to become square and resized to 256 for our experiment.

\subsection{Experiments}

We train our model separately on each dataset and analyse the image outputs both qualitatively and quantitatively. The qualitative evaluation involves a visual comparison of synthetic images produced by our approach against those generated by well-established baseline models, GD and SDM, to visually highlight the advancements our model offers in terms of image realism and fidelity.

\subsubsection{Metrics}

Quantitatively, we calculate standard metrics in image generation model evaluation: Fréchet Inception Distance (FID), Learned Perceptual Image Patch Similarity (LPIPS), Structural Similarity Index Measure (SSIM), and Peak Signal-to-Noise Ratio (PSNR). In measuring FID \cite{Heusel2017GANs} we choose to use a different feature layer of the Inception Network instead of the default \textit{pool3} layer. We use the \textit{first-max-pooling} and \textit{second-max-pooling} layers that capture fundamental image features without relying on the ImageNet-specific learned parameters, offering a more relevant evaluation for US images. The metrics of LPIPS, SSIM, and PSNR are computed for the CAMUS and Thyroid datasets, given their availability of semantic labels. These metrics are calculated for all corresponding pairs of images of the synthetical data and the original data. To compare the results and draw conclusions, we calculate the average, standard deviation, and range of these metrics.

\subsubsection{Training and Hyper-parameters}
We performed the training with a batch size of 4, 0.0001 learning rate and 2000 diffusion steps. The image resolutions varied: 128 for CAMUS and SegThy and 256 for Liver. The number of training iterations also differed, with 28,000 for CAMUS, 36,000 for Liver, and 50,000 for SegThy, stopping earlier for some datasets as they achieved convergence. Following previous works \cite{dhariwalDiffusionModelsBeat2021}, we selected a cosine scheduler for the $\alpha$ values and a square root scheduler for the B-maps. Through our experiments, we determined the optimal minimum values for the B-maps to be 0.96 and 0.97, which yielded the best results.

\section{Results and Discussion}

\begin{figure}[t]
    \centering
    \includegraphics[width=\textwidth]{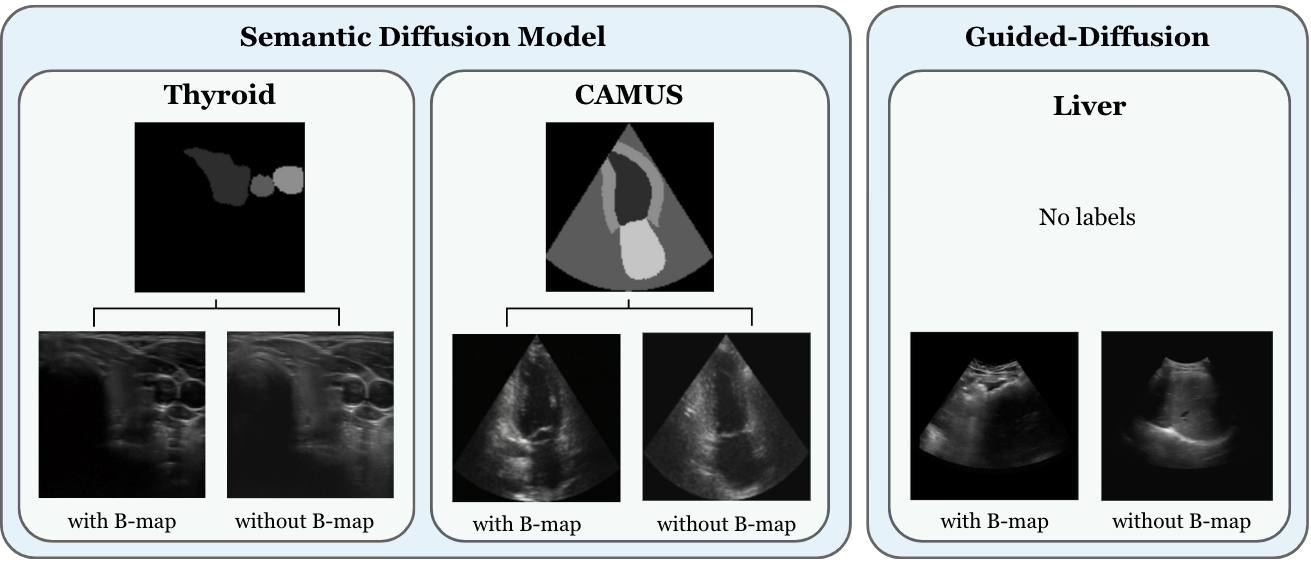}
    \caption{\textbf{Qualitative comparison}: The top row displays the label maps used for SegThy and CAMUS datasets. For the liver dataset, no labels were used. The bottom row shows the US images generated with B-Maps (left) versus without B-Maps (right) for each dataset.}
    \label{fig:qualitative}
\end{figure}

\textbf{Qualitative Analysis:} Visually, the integration of B-Maps has demonstrated a notable improvement in the synthetic images generated using both SDM and Guided-Diff. Figure \ref{fig:qualitative} showcases examples from each dataset, illustrating the improvement in image quality and realism, especially in terms of contrast in the upper regions of the images, aligning with the known characteristics of US scans. Additional visual comparisons are available in supplementary material (Sec.\ref{sec:supp}).

\noindent \textbf{Quantitative Evaluation:} The improvement achieved by B-Maps is also evident in the quantitative metrics used to evaluate image quality. FID scores, as shown in Table \ref{tab:fid_scores}, calculated with pytorch-fid \cite{Seitzer2020FID}, underscore the superiority of our proposed method over the baseline across all datasets. The scores from the 1\textsuperscript{st} and 2\textsuperscript{nd} max-pooling layers of the Inception Network for all synthetically generated images from the validation set demonstrate our method's ability to produce images that are closer to real US images, with a notable reduction in the FID scores. This improvement indicates that our generated images have higher fidelity and are statistically closer to the distribution of real US images. 

\begin{table}[h]
\centering
\caption{\textbf{FID Scores across Datasets.} Features for FID calculations are extracted from the \textit{first-max-pooling} (1\textsuperscript{st} MP) and \textit{second-max-pooling} (2\textsuperscript{nd} MP) layers of Inception. Results indicate that our method surpasses the baseline in generating datasets with more realistic images, as evidenced by the significantly lower FID scores.}
\label{tab:fid_scores}
\begin{tabular}{>{\raggedright\arraybackslash}p{0.3\textwidth} 
                >{\centering\arraybackslash}p{0.15\textwidth} 
                >{\centering\arraybackslash}p{0.15\textwidth} 
                >{\centering\arraybackslash}p{0.15\textwidth} 
                >{\centering\arraybackslash}p{0.15\textwidth}}
\toprule
 & \multicolumn{2}{c}{\textbf{Baseline}} & \multicolumn{2}{c}{\textbf{Proposed method}} \\
\cmidrule(l){2-5}
\textbf{Dataset} &  1\textsuperscript{st} MP & 2\textsuperscript{nd} MP & 1\textsuperscript{st} MP & 2\textsuperscript{nd} MP \\
\midrule
\textbf{Thyroid} & 4.259 & 14.315 & \textbf{0.619} & \textbf{2.769} \\
\textbf{CAMUS} & 3.581 & 12.769 & \textbf{0.204} & \textbf{0.959} \\
\textbf{Liver} & 20.746 & 73.447 & \textbf{0.192} & \textbf{0.867} \\
\bottomrule
\end{tabular}
\end{table}

In our evaluation, the LPIPS metric, in Table \ref{tab:comprehensive_analysis}, computed using PerceptualSimilarity \cite{zhang2018perceptual}, provides insight into the perceptual quality of generated images, reflecting how closely the synthetic data resembles real US scans. The results indicate that our method consistently yields lower LPIPS scores across Thyroid and CAMUS datasets, implying that images generated by our method are more realistic, aligning closely with the perceptual properties of real US scans.

Our SSIM and PSNR analyses, also detailed in Table \ref{tab:comprehensive_analysis} and computed using torchmetrics \cite{TorchMetrics2022},  offer additional insight into the image quality improvements achieved through our approach. While SSIM values show minimal differences from baseline methods—indicating comparable structural integrity—the PSNR values are significantly higher. This suggests that our method enhances image precision by improving the signal-to-noise ratio, thus generating clearer and sharper US images.

\begin{table}
\centering
\caption{\textbf{LPIPS, SSIM, and PSNR Metrics Comparison}: This table evaluates the quality of generated US images for Thyroid and CAMUS datasets using LPIPS, SSIM, and PSNR metrics. Lower LPIPS scores indicate closer resemblance to real images. Additionally, our approach achieves higher SSIM and PSNR values compared to baselines for both datasets, indicating better preservation of structural details and improved clarity, reflecting improved image quality.}
\label{tab:comprehensive_analysis}
\begin{tabular}{>{\raggedright\arraybackslash}p{0.08\textwidth} 
                >{\centering\arraybackslash}p{0.13\textwidth} 
                >{\centering\arraybackslash}p{0.13\textwidth} 
                >{\centering\arraybackslash}p{0.13\textwidth} 
                >{\centering\arraybackslash}p{0.13\textwidth} 
                >{\centering\arraybackslash}p{0.13\textwidth} 
                >{\centering\arraybackslash}p{0.13\textwidth}}
\cmidrule(l){1-6}
 & \multicolumn{3}{c}{\textbf{Baseline}} & \multicolumn{3}{c}{\textbf{Our Method}} \\
\cmidrule(l){3-6} 
\textbf{ } &  \textbf{Metric} &\textbf{Thyroid} & \textbf{CAMUS} & \textbf{Thyroid} & \textbf{CAMUS} \\
\cmidrule(l){1-6}
\multirow{3}{*}{\rotatebox[origin=c]{90}{\textbf{LPIPS}}} & Mean & 0.362 & 0.234 & \textbf{0.316} & \textbf{0.161} \\
& Std. Dev. & 0.129 & 0.127 & \textbf{0.066} & \textbf{0.047} \\
& Range & 0.628 & 0.534 & \textbf{0.321} & \textbf{0.266} \\
\cmidrule(l){1-6}
\multirow{3}{*}{\rotatebox[origin=c]{90}{\textbf{SSIM}}} & Mean & 0.279 & 0.265 & \textbf{0.292} & \textbf{0.297} \\
& Std. Dev. & 0.137 & 0.131 & \textbf{0.094} & \textbf{0.092} \\
& Range & 0.523 & 0.587 & \textbf{0.451} & \textbf{0.523} \\
\cmidrule(l){1-6}
\multirow{3}{*}{\rotatebox[origin=c]{90}{\textbf{PSNR}}} & Mean & 14.302 & 13.118 & \textbf{16.798} & \textbf{15.871} \\
& Std. Dev. & 5.526 & 4.511 & \textbf{2.468} & \textbf{2.545} \\
& Range & 19.707 & 18.899 & \textbf{11.549} & \textbf{10.420} \\
\cmidrule(l){1-6}
\end{tabular}
\end{table}

\section{Conclusion}
This study presents a novel approach to US image synthesis by adapting diffusion models with B-Maps. Our method introduces a customized noise schedule that reflects the natural attenuation of US waves. This innovation significantly enhances the realism of synthetic US images, as supported by our comprehensive evaluation across several datasets.

While our model showcases promising advancements in synthetic US generation, it also opens the door to exploring more sophisticated models that further incorporate ultrasound's physical properties. Future directions could involve developing models that estimate attenuation maps at each diffusion step, offering even more precise control over the synthetic image generation process.

\bibliographystyle{splncs04}
\bibliography{us_learner}

\footnote{The authors have no competing interests to declare that are relevant to the content of this article.}

\newpage

\pagenumbering{gobble}

\section*{Supplementary Material} \label{sec:supp}

\subsection*{Proof for the mathematical derivations for DDMPs with B-Maps}
\noindent For DDPMs with B-Maps implementation, like the standard DDMP, the optimization is based on maximizing the Evidence Lower Bound (ELBO)\cite{luoUnderstandingDiffusionModels2022}: 

\begin{equation}
\begin{split}
\log p(x) \geq  & \mathbb{E}_{q(x_1|x_0)}[\log p_\theta(x_0|x_1)] - D_{KL}(q(x_T|x_0) \| p(x_T)) \\
& - \sum_{t=2}^{T} \mathbb{E}_{q(x_t|x_0)}[D_{KL}(q(x_{t-1}|x_t, x_0) \| p_\theta(x_{t-1}|x_t))]
\end{split}
\end{equation}

In this derivation of the ELBO, the bulk of the optimization cost lies in the summation term. By Bayes rule, we have: $q(x_{t-1}|x_t, x_0) = \frac{q(x_t|x_{t-1}, x_0)q(x_{t-1}|x_0)}{q(x_t|x_0)} $

where the first term in the nominator is known Eq. (\ref{standard_forward}) and tractable. Recalling the reparameterization trick, the form of $q(x_t|x_0)$ and $q(x_{t-1}|x_0)$ can be recursively derived through repeated applications of the reparametrization trick. 
\begin{equation}
\begin{aligned}
    &x_t = \sqrt{\alpha_{t} B_t} x_{t-1} + \sqrt{1 - \alpha_{t}B_t}\epsilon_{t-1}, \quad \text{where } \epsilon_{t-1} \sim \mathcal{N}(0, \mathbf{I}) \\
    &= \sqrt{\alpha_{t}B_t} \sqrt{\alpha_{t-1}B_{t-1}} x_{t-2} + \underbrace{\sqrt{\alpha_{t}B_t} \sqrt{1 - (\alpha_{t-1}B_{t-1})}\epsilon_{t-2} + \sqrt{1 - (\alpha_tB_t)}\epsilon_{t-1}}_{(1)} \\
    &= \sqrt{\alpha_{t}\alpha_{t-1}B_tB_{t-1}} x_{t-2}+ \sqrt{1 - (\alpha_t\alpha_{t-1}B_tB_{t-1})}\epsilon_{t-2} \\
    &= \ldots \\
    &= \sqrt{\prod_{i=1}^{t} \alpha_iB_i} x_0 + \sqrt{1 - \prod_{i=1}^{t} \alpha_iB_i} \epsilon, \\
    &\sim \mathcal{N}(x_t; \sqrt{\bar{\alpha}_t\bar{B}_t} x_0, (\mathbf{1} - \bar{\alpha}_t\bar{B}_t)
\end{aligned}
\label{eq:refined_diffusion_process}
\end{equation}
Where, all the multiplications are point-wise and in (1) we apply the fact that the sum of two independent Gaussian random variables remains a Gaussian with mean being the sum of the two means, and variance being the sum of the two variances. We have therefore derived like this the Gaussian form of $q(x_t|x_0)$, and we can derive $q(x_{t-1}|x_0)$ the same way. Therefore, knowing the forms of both $q(x_t|x_0)$ and $q(x_{t_1}|x_0)$, we can proceed to calculate the form of $q(x_{t_1}|x_t, x_0)$:

\begin{multline}
q(x_{t-1}|x_t, x_0) = \frac{q(x_t|x_{t-1}, x_0)q(x_{t-1}|x_0)}{q(x_t|x_0)} \\
= \frac{\mathcal{N}(x_t; \sqrt{\alpha_t B_t}x_{t-1}, (\mathbf{1} - \alpha_t B_t)\mathbf{I})\mathcal{N}(x_{t-1}; \sqrt{\bar{\alpha}_{t-1}\bar{B}_{t-1}}x_0, (\mathbf{1} - \bar{\alpha}_{t-1}\bar{B}_{t-1})\mathbf{I})}{\mathcal{N}(x_t; \sqrt{\bar{\alpha}_t \bar{B}_t} x_0, (\mathbf{1} - \bar{\alpha}_t \bar{B}_t) \mathbf{I})}
\end{multline}

Following the same derivation as in \cite{luoUnderstandingDiffusionModels2022} but with our modified Gaussian distributions that depend also on $B$, we arrive to the posterior distribution described in the paper. 
\vspace{-100pt}
\begin{figure}[ht]
    \centering
    \includegraphics[width=\textwidth]{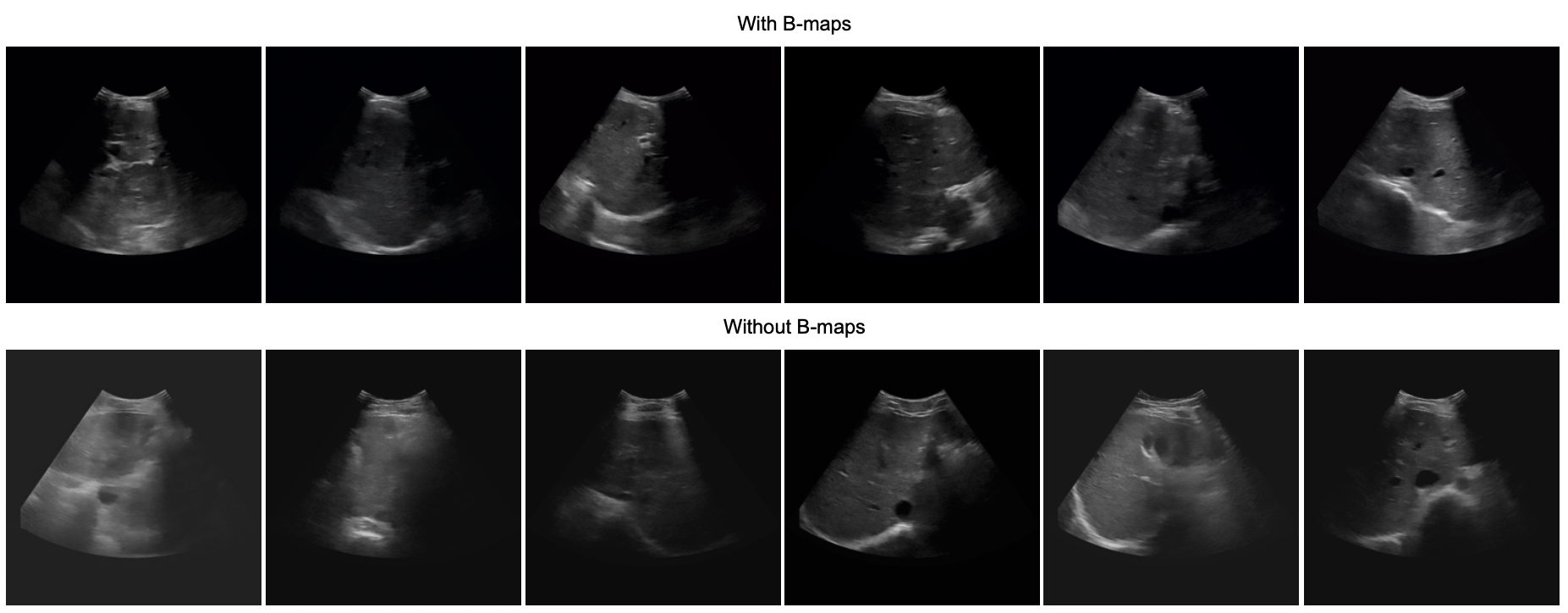}
    \caption{\footnotesize\textbf{Liver Results}: US images generated with B-Maps (top) exhibit enhanced contrast, especially in the upper regions, compared to those without (bottom).}
\end{figure}
\vspace{-100pt}
\begin{figure}[ht]
    \centering
    \includegraphics[width=\textwidth]{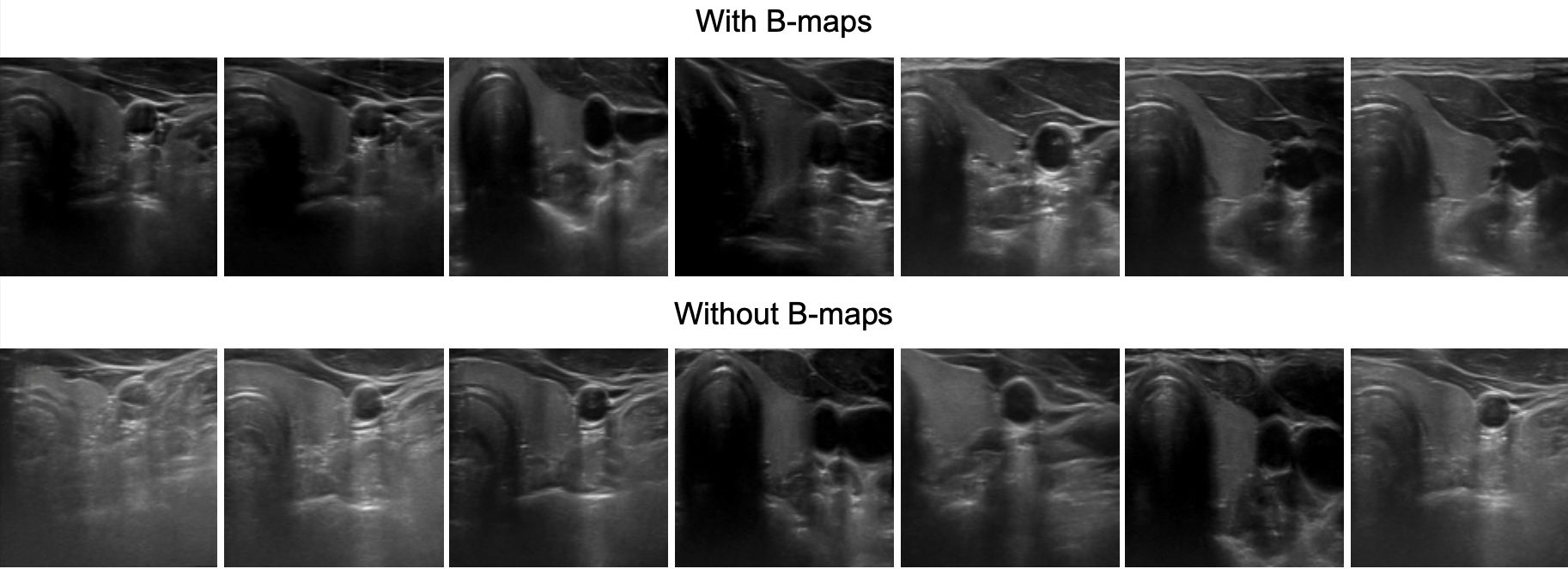}
    \caption{\footnotesize\textbf{Thyroid Results}: The application of B-Maps in the generation of thyroid US images (top row) results in a clearer delineation and contrast, which is less pronounced in images generated without B-Maps (bottom row).}
\end{figure}
\vspace{-100pt}
\begin{figure}[ht]
    \centering
    \includegraphics[width=\textwidth]{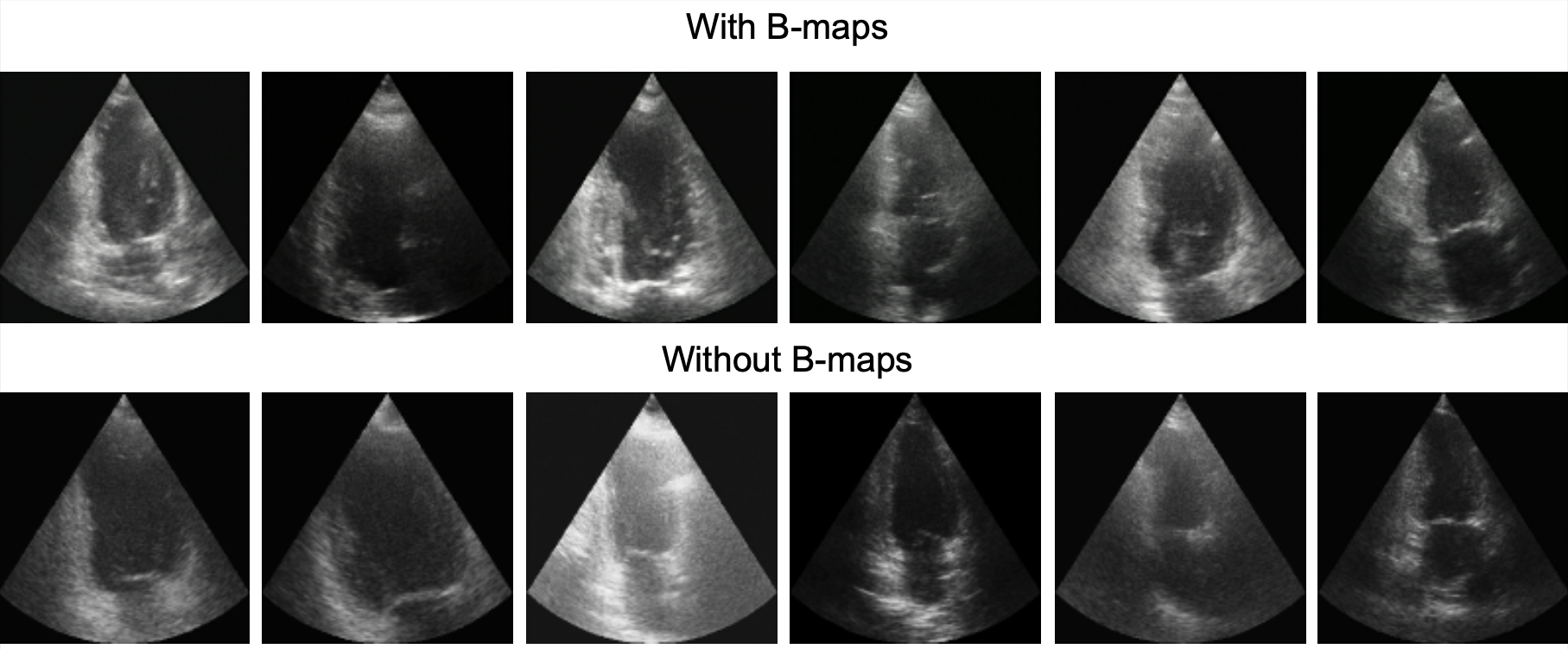}
    \caption{\footnotesize\textbf{CAMUS Results}: Images generated with B-Maps (top row) have improved contrast in the superior sections, contrasting with the lower contrast seen in the images without B-Maps (bottom row).}
\end{figure}

\end{document}